\documentclass[10pt,onecolumn]{elsart3p}
\usepackage{graphicx}
\usepackage{amsmath}
\usepackage{amssymb}
\journal{Optics Communications; Accepted 5 November 2008; www.elsevier.com}

\begin{document}

\begin{frontmatter}

\title{Between Right and Left Handed Media}
\author{P C Ingrey\corauthref{cor}}
\corauth[cor]{Author to whom any correspondence should be addressed.}
\ead{pmxpi@nottingham.ac.uk}
\author{K I Hopcraft}
\author{E Jakeman}
\author{O E French}

\address{School of Mathematical Sciences, University Park, University of Nottingham, NG7 2RD, England}

\begin{abstract}
We report analytical calculations for the propagation of electromagnetic radiation through an inhomogeneous layer whose refractive index varies in one dimension situated between bulk right- and left-handed media.  Significant field localization is generated in the layer that is caused by the coherent superposition of evanescent waves. The strength of the field localization and the transmission properties of the layer are investigated as a function of the layer width, losses and defects in the refractive index; the former two being modelled by continuous changes, and the latter by discontinuous changes, in the index profile.
\end{abstract}

\end{frontmatter}


\section{Introduction}
\label{Cp:Intro}
Metamaterials are of importance because of their ability to manipulate the behaviour of electromagnetic radiation in ways that cannot be achieved by naturally occurring matter \cite{Cloak,MCloak,Spliter}. They function through fabricating the microscopic electromagnetic properties to create an effective medium, and Maxwell's macroscopic equations provide the description of the fields therein. A class of metamaterials with permittivity and permeability simultaneously less than zero have been fabricated \cite{SmithCreation,VisLHM,ThreeDOptical}, these having been postulated earlier \cite{Ves}. These ‘left handed media’ (LHM) are so called because their $\bold{E}$, $\bold{H}$ and $\bold{k}$ vectors form a left-handed co-ordinate set with the consequence that their wave propagation vector is anti-parallel to the direction of energy flow. The unusual optical properties of such materials have stimulated great interest, not least because of their ability to create lenses that are not diffraction limited, and for cloaking \cite{Cloak,MCloak,PerfectLens}, whilst providing a striking example of resonant behaviour in physics \cite{PerfectLens}. However, the resonant behaviour implies that the interaction of radiation with LHM is a singular phenomenon and is therefore susceptible to imperfections in the material parameters and the way that they are configured in bulk matter. 

Left handed media are necessarily dispersive \cite{PendrysRetortToRetort} which leads to losses that affect substantially the efficiency of components constructed from them. The values of the material parameters required for their applications to operate successfully are exacting \cite{PerfectButExact,OllieOffMinusOne}, and deformations and contamination of the idealized planar surfaces between components can also severely limit performance  \cite{OllieMieScattering,TransSlab}. This paper examines the effect of an imperfect boundary between bulk right- and left-handed media, such as may be caused by a slight roughening of the interface, or due to homogenization of the material parameters through the intermingling of bulk material of different effective refractive index.

The model adopted will assume that the thickness of the transition layer between the bulk right- and left-handed material is small compared with the smallest lateral scale-size that characterizes the interface. Consequently the change in refractive index can be assumed to be limited to one-dimension. Maxwell's equations then reduce to an ordinary differential equation, for which the solution is expressible in terms of exponential functions in the bulk media, and in terms of special functions for particular profiles of the refractive index within the layer. In particular a linear transition between refractive index of +1 and -1 has an analytical solution within the layer in terms of confluent hypergeometric functions, and it is the properties of this exact solution that are explored in this paper. 

The following section formulates the model for the layer and discusses the general forms of the solution to Maxwell's equations. Section \ref{cp:lossless} obtains the solution for the linear refractive index profile and discusses both the form and interpretation of the solution in the absence of losses in the material parameters. The most striking feature of this solution is a localization of the field within the layer. Section \ref{cp:lossy} reconsiders the model in the presence of finite losses and shows that the lossless solution is singular in several respects, not least that losses destroy the localization and perfect transmission property of the layer. Section \ref{cp:delta} shows that a small discontinuity in the refractive index profile obtains solutions whose properties are intermediate between the lossless and lossy cases. The concluding section summarizes and discusses the results.

\section{The GRIN model}
\label{cp:grin}
Graded-index (GRIN) modelling of a boundary between bulk media has been used in conventional RHM for many years \cite{Lekner} and is tantamount to considering the properties of waves that interact with a one-dimensional refractive index profile
\begin{equation} \label{eq:GRINn}
n(z) = \left\lbrace 
\begin{array}{ll}
n_1 & z < z_1 \\
\nu (z) & z_1 < z < z_2 \\
n_2 & z_2 < z
\end{array} \right. ,
\end{equation}
\begin{figure}[t!]
\center
\includegraphics[width=0.35\textwidth]{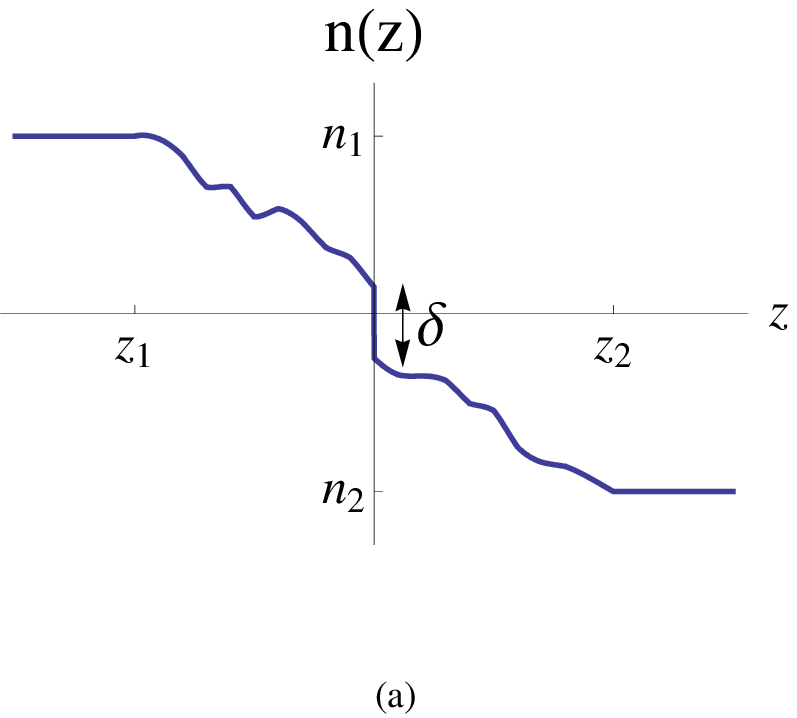}
\includegraphics[width=0.4\textwidth]{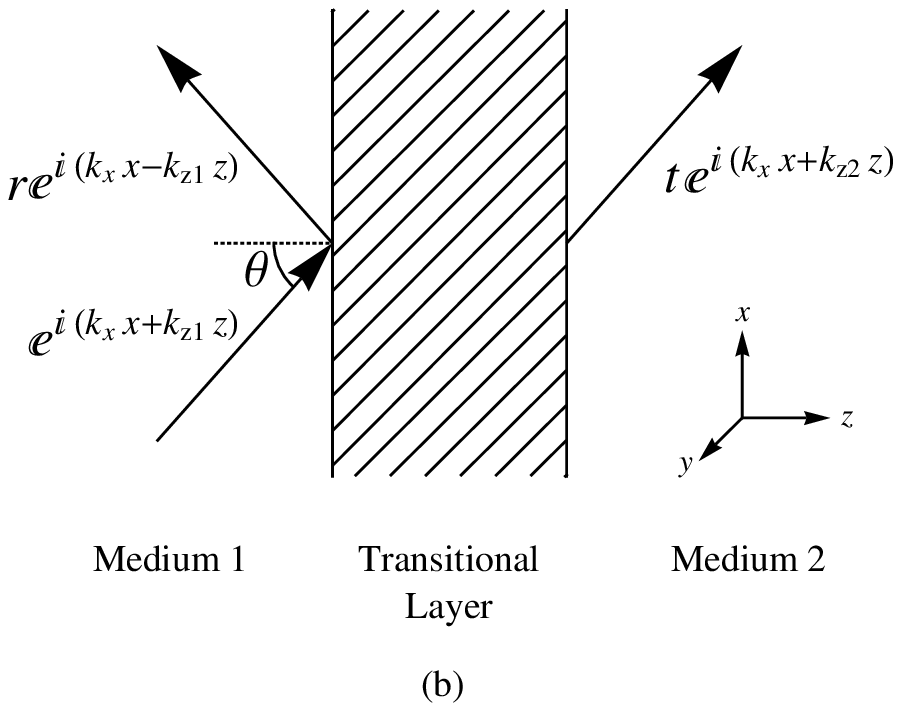}
\caption{\small{(Colour Online) (a) A typical one-dimensional GRIN layer profile containing a step discontinuity of $\delta$. (b) Pictorial representation of the incoming planar wave and the resulting reflected and transmitted waves.}}
\normalsize
\label{fig:GRIN}
\end{figure}
where $\nu(z_1) = n_1$, $\nu(z_2) = n_2$ with $n_1$ and $n_2$ being the bulk refractive indices in media 1 and 2 respectively\footnote{More specifically $\epsilon(z_1) = \epsilon_1$, $\epsilon(z_2) = \epsilon_2$, $\mu(z_1) = \mu_1$ and $\mu(z_2) = \mu_2$ with subscript 1 indicating medium 1 and subscript 2, medium 2}, The refractive index in the layer can be continuous, piecewise continuous or contain discontinuities of size $\delta$, as shown in figure \ref{fig:GRIN} (a). There is little restriction on the profile of $\nu(z)$, but only a few analytical solutions of Maxwell's equations with profile (\ref{eq:GRINn}) exist for problems involving LHM.

A harmonic $s$-polarized electromagnetic wave travels in the direction of increasing $z$ through a lossless right handed medium of refractive index $n_1$, and at angle $\theta$ to the normal of the transitional layer, as shown in figure \ref{fig:GRIN} (b). The form of the electric field is $\bold{E}(x,y,z) = E_y (x,y,z) \, \bold{\hat{y}}$ (a harmonic time dependence, $\omega$ is assumed throughout). Because the refractive index is a function of $z$ only, Maxwell's equations lead to separable solutions of the form $E_y (x,y,z) = \exp(i k_x x) \, E(z)$, where $k_x$ is the wave number in the $x$ direction which is in the plane of the interface between the bulk media. In the homogeneous bulk media the form adopted by $E(z)$ are simple exponential functions, whereas in the layer the field is a solution of the equation
\begin{equation}\label{eq:ODE}
\frac{d}{d z} \left( \frac{1}{\mu(z)} \frac{d E(z)}{d z} \right) + \left( \frac{ \omega ^2 \epsilon(z)}{c^2} - \frac{k_x^2}{\mu(z)} \right) E(z) = 0
\end{equation}
which results in the electric field adopting the form
\begin{equation} \label{eq:GRINElec}
E(z) = \left\lbrace 
\begin{array}{ll}
\exp(i \, k_{z_1} \, z) + r \exp(-i \, k_{z_1} \, z) & z < z_1 \\
\tilde{E}(z) & z_1 < z < z_2 \\
t \exp(i \, k_{z_2} \, z) & z_2 < z
\end{array} \right. ,
\end{equation}
where $\tilde{E}$ is the electric field in the layer, $c$ is the speed of light in vacuum, $k_{z_1}$ and $k_{z_2}$ are the wave number in the $z$ direction in media 1 and 2 respectively. The reflection and transmission coefficients $r$ and $t$ are determined once the form for $\tilde{E}$ is known by invoking the continuity of the tangential components of $\bold{E}$ and $\bold{H}$ at both extremities of the layer at $z_1$ and $z_2$.

A number of profiles for $\epsilon(z)$ and $\mu(z)$ give rise to analytic solutions to equation (\ref{eq:ODE}). Of these, there are two that can traverse between RHM and LHM: that of an exponential profile and that of a linear profile. Only the linear profile allows for independent choice of the losses within the model, and therefore will be adopted. The results of the exponential case are given in Appendix \ref{cp:exp}. The profiles for $\mu(z)$ and $\epsilon(z)$ for the case of interest are taken to be
\begin{equation} \label{eq:slprofile}
\mu(z) = m \, z + d , \quad \epsilon(z) = \eta \, \mu(z) .
\end{equation}
It is clear from these definitions that with complex $m$ and $d$, this profile can be used to model the transition between two bulk media of any $\mu$, whether lossy or not. The limitation that $\epsilon$ cannot be chosen independently of $\mu$ is not especially restrictive given that most devices which employ negative refraction or perfect lensing have $\epsilon$ and $\mu$  everywhere equal \cite{Cloak,PerfectLens}, i.e. that $\eta$ is unity in equation (\ref{eq:slprofile}). Consequently $\eta$ will be set to unity in the remaining sections.

For the profile given by (\ref{eq:slprofile}) the electric field within the layer is given by
\small
\begin{equation} \label{eq:slelec}
E(z) = \frac{\exp \left( \frac{- i \, \gamma\, \Psi(z)}{2} \right) \Psi(z) }{4\,c^2 \,m^2} \left( \alpha F(z) + \beta G(z) \right)
\end{equation}
\normalsize
where
\small
\begin{equation*}
F(z) = \mbox{M} \! \left( 1 - \frac{i k_x^2}{4\,m^2\,\gamma},2, i \, \gamma \, \Psi(z) \right) ,
\end{equation*}
\begin{equation*}
G(z) = \mbox{U} \! \left( 1 - \frac{i k_x^2}{4\,m^2\,\gamma},2,i \, \gamma \, \Psi(z) \right) ,
\end{equation*}
\begin{equation*}
\gamma = \frac{\eta^{1/2} \,{\omega }}{c\,m} , \quad \Psi(z) = { \left( d + m\,z \right) }^2 = \mu^2,
\end{equation*}
\normalsize
$M$ and $U$ are the independent confluent hypergeometric functions \cite{Handbook} and $\alpha$ and $\beta$ are constants of integration. Using the conditions that
\begin{equation} \label{eq:BC}
E(z) \quad \mbox{and} \quad \frac{1}{\mu(z)} \frac{d E(z)}{d z}
\end{equation}
are continuous across both boundaries between the inhomogeneous layer and the bulk media uniquely determines the four unknowns, $r$, $t$, $\alpha$ and $\beta$. These have a similar structure to those given in \cite{Lekner}. In the first instance this will be used to determine the electric field throughout a diffuse boundary between lossless RHM and lossless LHM.

\begin{figure}[!t]
\center
\includegraphics[width=0.4\textwidth]{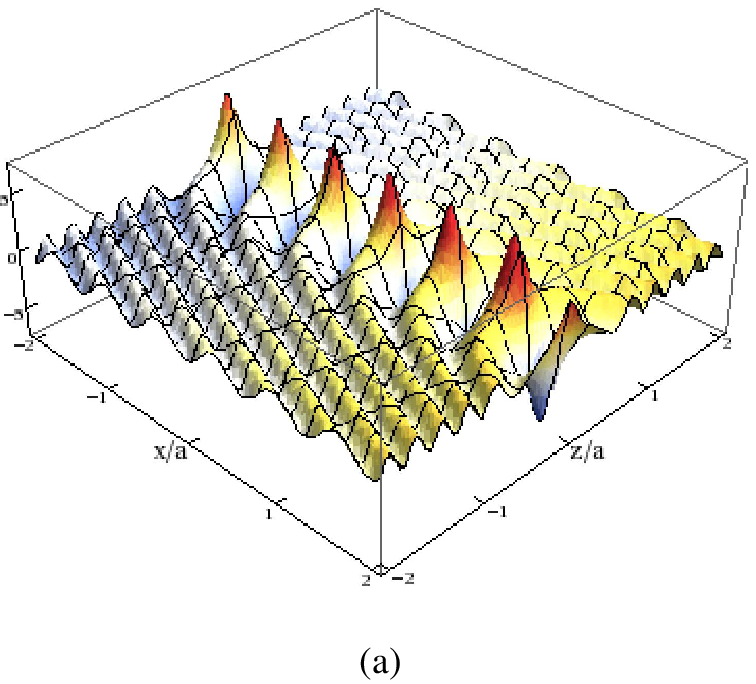}
\includegraphics[width=0.35\textwidth]{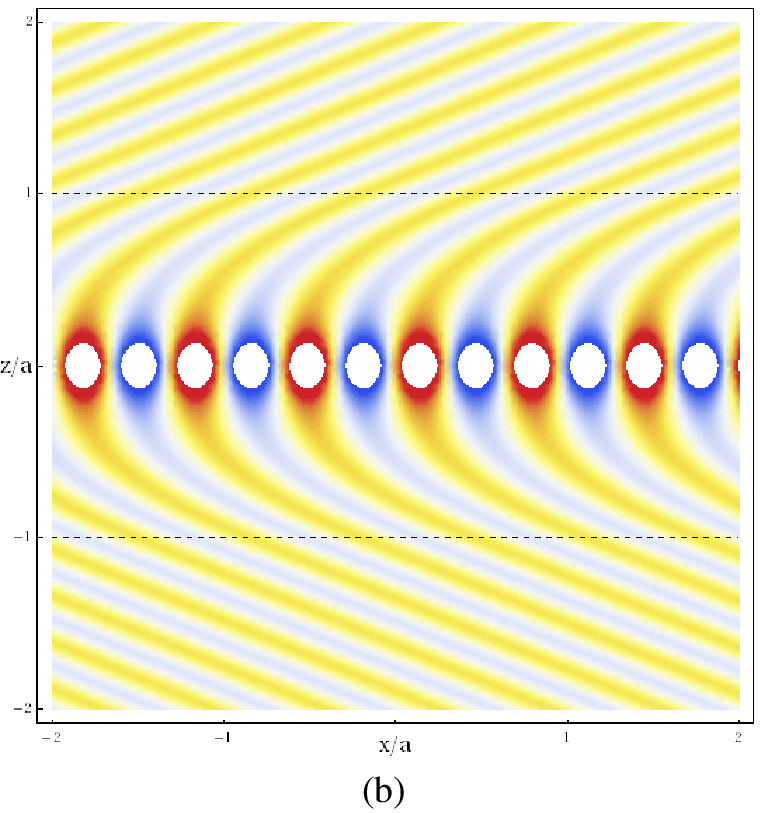}

\includegraphics[width=0.4\textwidth]{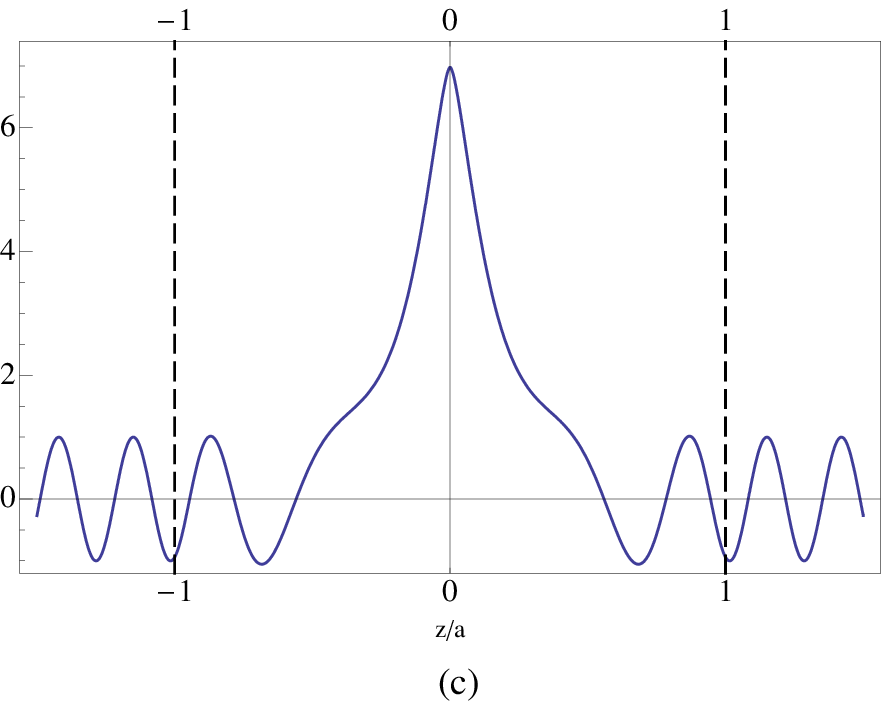}
\caption{(Colour Online) The real part of $E$ that results for a dimensionless layer width, $k a = 8 \pi$ for an angle of incidence $\theta=\pi/8$ to the normal of the surface from a vacuum ($z < -1$) to a bulk medium with $\epsilon = \mu = -1$ ($z > 1$). (a) shows a 3D representation of the real part of the electric field, note the localizations move with speed $\omega / k_x$ in the $x$ direction, (b) the same situation in contour form and (c) a cross section of part (a) in the constant $x$ plane. The amplitude of the field is taken to be unity in the homogeneous media.}
\label{fig:lossless}
\end{figure}

\section{The lossless solution} \label{cp:lossless}
\label{cp:RightToLeft}
Without loss of generality, the layer is assumed to be of width $2a$ and is centred at the origin. Figure \ref{fig:lossless} shows various depictions of the real part of the electric field when the $s$-polarized wave is incident from a right-handed medium ($n=1$) into a left-handed media ($n=-1$) at an angle of incidence $\theta=\pi/8$, and when the dimensionless layer width $ka = 8\pi$. 

The surface plot figure \ref{fig:lossless} (a) illustrates the localized field confined within the transition layer. Figure \ref{fig:lossless} (b) depicts the level contours of the field and clearly shows that negative refraction has occurred from the orientation of the phase fronts far from the layer. Figure \ref{fig:lossless} (c) is a cross-section across the layer, and shows that the field within the layer is enhanced several times in excess of its (arbitrary) value in the homogeneous media. The localized structure is, moreover, symmetrical about the origin, there being no way that an asymmetrical solution can match onto the field structures in the bulk media.

Although the refractive index changes smoothly from +$1$ to $-1$ with increasing $z$, the impedance $Z=1$ everywhere, and so the layer is perfectly transparent to the radiation, i.e. $t =1$ and $r =0$. Consequently if the fields were observed in the bulk media alone, it would not be possible to infer the presence or absence of the layer.

The origin of the localized field can be understood with reference to figure 3, which is meant for illustrative purposes alone and which depicts the field through the layer, within which $\left| n \right| <1$. In the region $-1<z/a<0$, both $\epsilon$ and $\mu$ are positive and so when the value of $\left| n \right|$ falls below that which allows a propagating mode to exist, i.e. $\left| n(z) \right| < \sin \theta$, an evanescent mode is established that decays with increasing $z$. This evanescent mode has a finite amplitude at $z =0$. For $0<z/a<1$, both $\epsilon$ and $\mu$ are negative, with result that the evanescent wave is amplified out to a distance $z_c$ satisfied by $\left| n(z_c) \right| = \sin \theta$, whereupon the wave can propagate once again for $z>z_c$. An evanescent wave is reflected back into the layer that \emph{increases} with decreasing $z$, until $z<0$, where the mode then decreases in the right-handed medium. Thus a coherent structure is established in the region $\left| n(z) \right|<z_c$ contained within the layer through the interference of these evanescent modes. This prompts questioning how the peak magnitude of the field scales with the layer width. This dependence is shown in figure \ref{fig:lossy} (a) by the red (triangles) curve, and shows an exponential growth with the layer thickness. This dependence obtains from the complicated nature of the confluent hypergeometric functions and their derivatives, that are contained within $\alpha$ and $\beta$. Clearly this behaviour is unphysical and will be ameliorated by the incorporation of losses, as will be considered in the next section. The red (triangles) curve in figure \ref{fig:lossy}(b) shows the peak value of the field as a function of the angle of incidence for a dimensionless layer width $ka = 8\pi$. This increases with increasing $\theta$ since the optical distance through the amplifying layer increases with increasing layer width.
\begin{figure}[t!]
\center
\includegraphics[width=\textwidth]{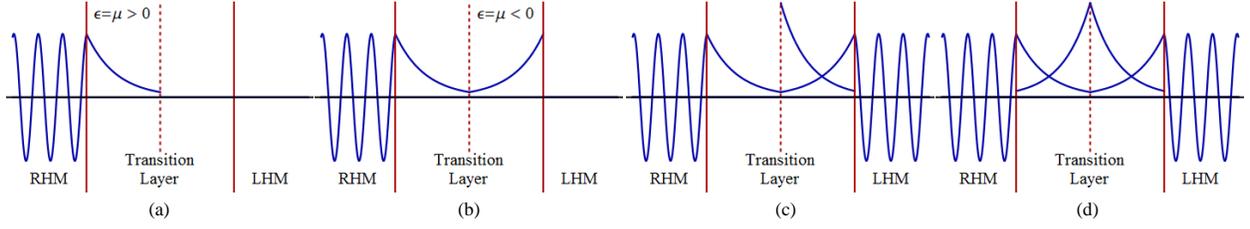}
\caption{\small{(Colour Online) Within the inhomogeneity layer there is a band where an incoming wave becomes evanescent in nature, $\left| n \right|$ being too low to allow propagation (a). These evanescent modes decay within the RHM and exponentially grow in the LHM \cite{PerfectLens} (b). When an evanescent wave reaches a point where it can propagate again the boundary conditions cause an evanescent wave to be reflected back into the layer (c). This propagates to the other side where another reflection occurs (d). It is the summation of each of these evanescent reflections that gives rise to the structure seen in figure \ref{fig:lossless} (c).}}
\normalsize
\label{fig:evalayer}
\end{figure}

\section{Lossy Transition From $n = 1$ to $n = -1 + \kappa i$} \label{cp:lossy}
In this section losses are incorporated into the bulk left-handed medium, which is assumed to have refractive index $n = -1 + \kappa i$ with $\kappa > 0$. The inhomogeneous layer is also assumed to be lossy with profile:
\begin{equation}
\mu(z) = \epsilon(z) = \left( -1 + \frac{\kappa}{2} i \right) \frac{z}{a} + \frac{\kappa}{2} i
\end{equation}

\begin{figure}[t!]
\center
\includegraphics[width=0.4\textwidth]{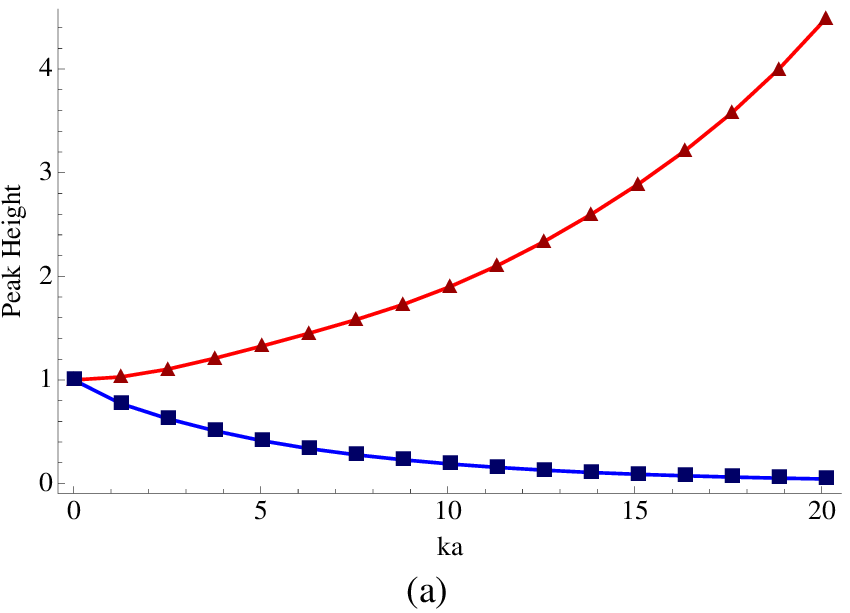}
\includegraphics[width=0.4\textwidth]{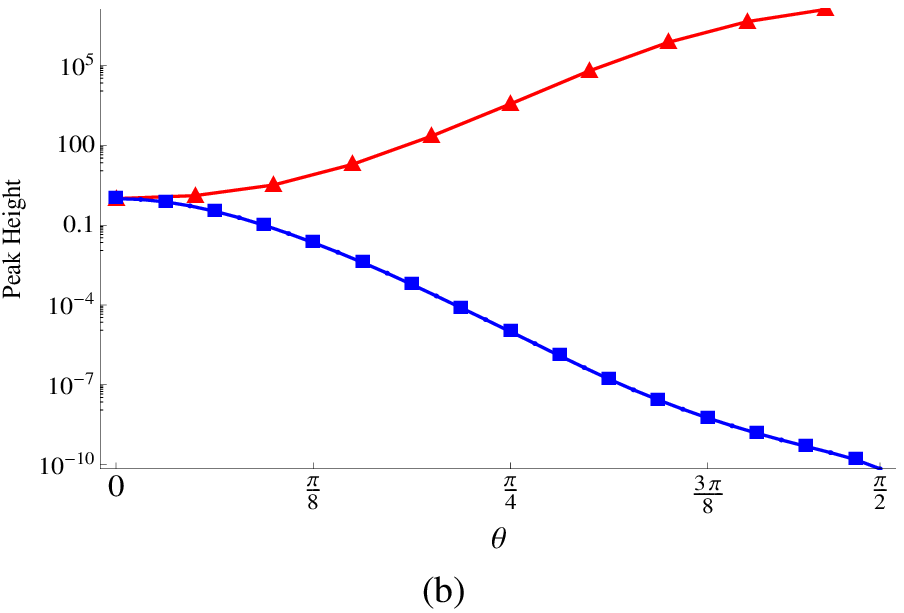}
\caption{\small{(Colour Online) A wave of wave number $k$, propagating at an angle $\theta$ to the normal, transitions from a vacuum through the inhomogeneity region, of width $2 a$, into a medium where  $\mu = -1 + \kappa i$. (a) shows the magnitude of the localisation in the medium relative to the incident wave with $\theta = \pi/8$ for $\kappa = 0$ (lossless) [Red Triangles] and $\kappa = 10^{-3}$ [Blue Squares]. (b) also shows the magnitude of the localisation but as a function of the incident angle, $\theta$, for no losses [Red Triangles] and $\kappa = 10^{-5}$ [Blue Squares]}}
\normalsize
\label{fig:lossy}
\end{figure}

According to this model, the refractive index is purely imaginary at the origin and the material is no longer impedance matched for all values of $z$. These two properties imply that the field will no longer grow without bound with increasing layer width and that the layer is no longer perfectly transparent to the radiation, i.e. a reflected wave exists in the right-handed half-space, $z<0$. Indeed the presence of losses implies that for $z \gg 0$, the wave will have decayed completely. The structure of the solution within the layer is also substantially modified by the losses, for the second confluent hypergeometric function possesses a branch cut which must be crossed as the layer is traversed. It can be shown that for a change in refractive index from $n_1 = \phi_1 + \kappa_1 i$ at $z_1$ to $n_2 = \phi_2 + \kappa_2 i$ at $z_2$ then the position $\overline{z}$ where the branch cut is crossed is given by:
\begin{equation} \label{eq:fullform}
\overline{z} = \frac{\left(\kappa_1 \, \phi_2 -  \kappa_2 \, \phi_1\right) \left| \Delta z \right| + \left( \Delta \kappa^2 + \Delta \phi^2 \right)^{1/2} \left(z_1 \, \phi_2 -  z_2 \, \phi_1\right) }{\left( \Delta \kappa^2 + \Delta \phi^2 \right)^{1/2} \Delta \phi}
\end{equation}
with $\Delta \psi = \psi_2 - \psi_1$ for any $\psi$. For $z > \overline{z}$, U can be calculated through the use of

\begin{equation}
\mbox{U} \! \left(a, \, 2, \, Z(z) \right) =  \overline{\mbox{U}} \! \left(a, \, 2, \, Z(z) \right) - \, \mbox{H}\! \left( z - \overline{z} \right) \frac{ 2 \, \pi \, i \, \, \mbox{Sign}  \left[ \mathbb{R}e \left( \left. n  \right|_{z=\overline{z}} \right) \right]}{\Gamma \left( a - 1\right)} \mbox{M} \! \left(a, \, 2, \, Z(z) \right) 
\end{equation}
where $\overline{z}$ is given in (\ref{eq:fullform}), $a$ and $Z(z)$ are the first and third argument of the U found in equation (\ref{eq:slelec}), $\mbox{H}$ is the Heaviside step function and $\overline{\mbox{U}}$ is the principal branch of the hypergeometric function \cite{Handbook} that is defined when $ -\pi< \arg{Z(z)} \leq \pi $. This allows for calculations involving losses to be performed in a similar way to the previous sections. Figure \ref{fig:lossy} details a variety of different features resulting from the addition of losses. The main point to note is that figure \ref{fig:lossy} (a) shows the exponential dependence of the localisations on layer width is suppressed by the addition of losses.

\begin{figure}[t!]
\center
\includegraphics[width=0.4\textwidth]{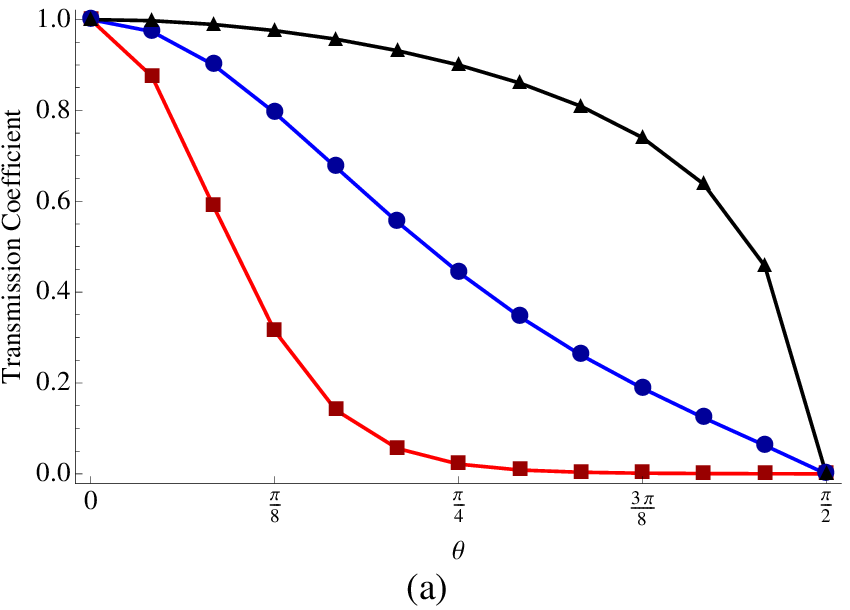}
\includegraphics[width=0.4\textwidth]{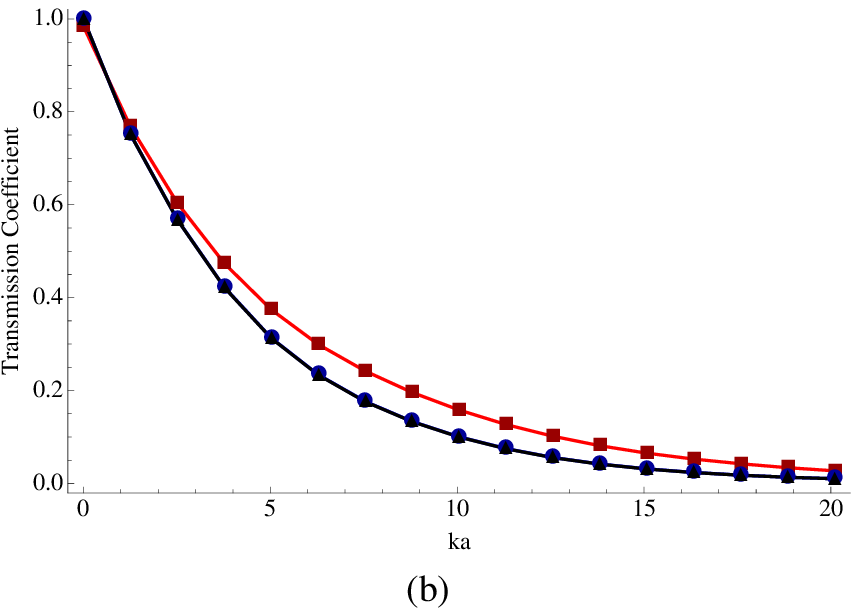}
\caption{\small{(Colour Online) For the situation detailed in figure \ref{fig:lossy}, (a) displays the dependence of the transmission coefficient on $\theta$ for various $ka$, this is obtained with $\kappa = 10^{-3}$ for  $ka = 5$ [Red Squares], 1 [Blue Circles] and 0.1 [Black Triangles] and (b) shows the transmission coefficient as a function of $ka$ at $\theta = \pi/8$ for $\kappa = 10^{-1}$ [Red Squares], $10^{-3}$ [Blue Circles] and $10^{-10}$ [Black Triangles] (The latter two are indistinguishable)}}
\normalsize
\label{fig:lossy2}
\end{figure}

A careful treatment of the branch-cut shows that the form of the solution is affected. The blue (squares) curve in figure \ref{fig:lossy} (a) shows that the peak value of the field in the layer now decreases exponentially with increasing layer width for a modest value of $\kappa = 10^{-3}$, as is also the case for the variation with $\theta$ as shown in figure \ref{fig:lossy} (b). Indeed the peak value of field within the layer is less than that outside it, so that the localization effect is entirely suppressed. This is because the evanescent modes are dissipated by the losses. Despite these losses, the field within the layer is still finite once the location where the modes can propagate again is attained, and so there is still transmission of radiation into the bulk left-handed medium. Figure \ref{fig:lossy2} (a) shows the transmission coefficient at $z = a$ for $\kappa = 10^{-3}$ as a function of the angle of incidence for a selection of values of the dimensionless layer width, $ka=0.1$ (brown triangles), $ka=1$ (blue circles) and $ka=5$ (red squares). The layer becomes opaque to radiation at progressively smaller angles of incidence as the layer thickness increases.  Figure \ref{fig:lossy2} (b) shows the negative exponential dependence of the transmission coefficient as a function of $ka$ for different values of the loss, $\kappa = 10^{-1}$ (red squares),  $\kappa = 10^{-3}$ (blue circles) and  $\kappa = 10^{-10}$ (blue triangles), the latter two being essentially indistinguishable. Hence for losses of the order $10^{-3}$ or lower the transmission coefficient is insensitive to the precise value of $\kappa$: this is an artefact of the dampening of the evanescent reflections within the transition layer. It can also be noted that the transmission coefficient, for fixed $ka$, increases as the losses increase. This effect occurs because large losses quickly damp all but the first reflection of the evanescent mode within the layer which leads to a larger ratio between the transmitted and reflected wave. The total transmitted and reflected power, however, steadily decrease as losses increase.

\section{The effect of discontinuities in the refractive index profile - the $\delta$-GRIN model } \label{cp:delta}
This previous section showed that the lossless results are very different from those where losses are included and this is principally because the refractive index vanishes at $z=0$ in the former case, whereas there is a branch-cut in the solution for the latter. This section will examine the robustness of these two classes of solution by modelling the refractive index in the layer by a ‘staircase’. Within each plateau of the staircase, the refractive index is constant and so the solution to Maxwell's equations is comprised of two independent exponentials. Using the boundary conditions (\ref{eq:BC})  at the end of each of the $N$ steps gives $2N$ equations for the $2N$ constants that determine the amplitudes and phases throughout the layer \cite{Lekner}.

\begin{figure}[t!]
\center
\includegraphics[width=0.7\textwidth]{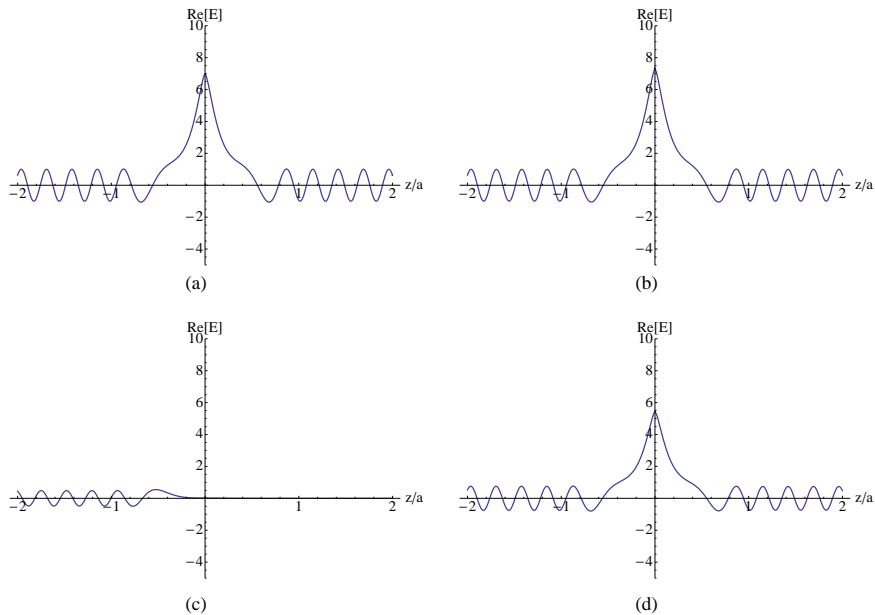}
\caption{\small{(Colour Online) (a) A typical electric field for a lossless GRIN, (b) for a lossless staircase with $N= 2^7$, (c) for a lossy ($\kappa = 10^{-5}$) GRIN and (d) for a lossy staircase with $N= 2^7$.}}
\normalsize
\label{fig:GRINvsStair}
\end{figure}

\begin{figure}[t!]
\center
\includegraphics[width=0.4\textwidth]{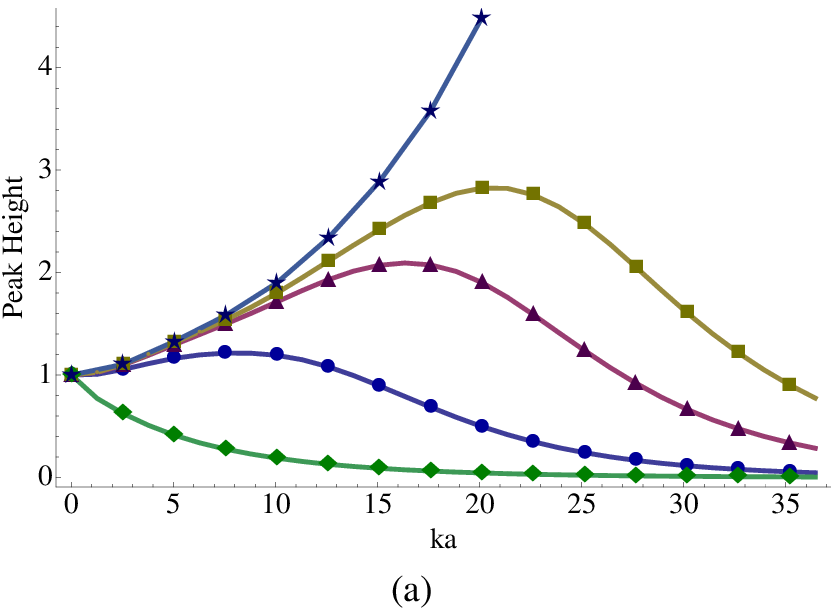}
\includegraphics[width=0.4\textwidth]{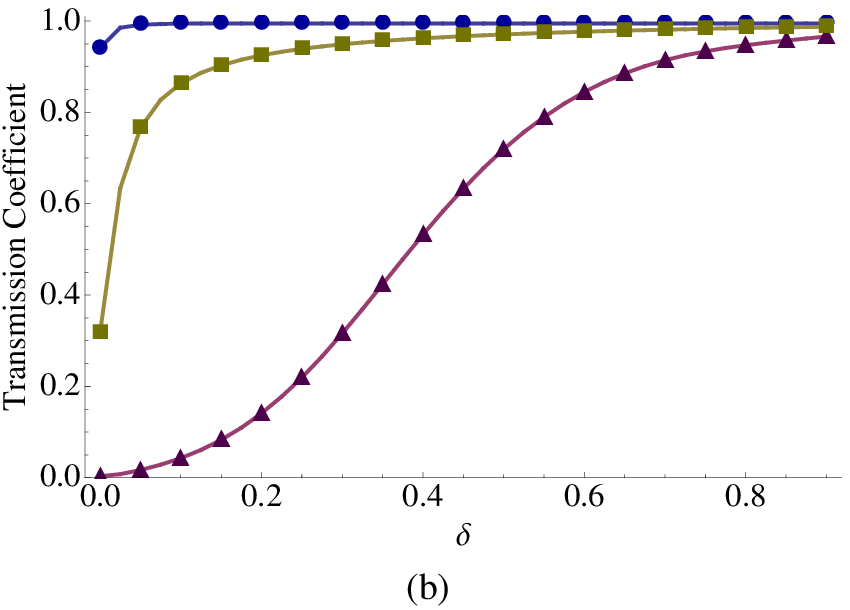}
\caption{\small{(Colour Online) (a) Localisation height with the parameter values of figure \ref{fig:lossy} (a) but with the addition of a discontinuity, $\delta$. The case shown are $\kappa = 10^{-3}$ and $\delta = 0$ [Green Diamonds], $\kappa = 10^{-3}$ with $\delta$ = 0.01 [Blue Circles], $\kappa = 10^{-3}$ with $\delta$ = 0.05 [Purple Triangles], $\kappa = 10^{-3}$ with $\delta$ = 0.1 [Dark Yellow Squares] and $\kappa = 0$ with no $\delta$ [Light Blue Stars]. (b) Transmission as a function of $\delta$ for $ka = 1$ [Purple Triangles] , $ka = 10^{-1}$  [Dark Yellow Squares] and $ka = 10^{-2}$ [Blue Circles].}}
\normalsize
\label{fig:PHwdelta}
\end{figure}

Figure \ref{fig:GRINvsStair} contrasts the solutions obtained from the continuum models of sections \ref{cp:grin}-\ref{cp:lossy} and the discrete staircase model. Figure \ref{fig:GRINvsStair}(a) is for a lossless medium and is repeated from figure \ref{fig:lossless}(c) for ease of comparison; this should be compared with \ref{fig:GRINvsStair}(b), which is for a staircase with $N=2^7$ equally spaced plateaus throughout the layer, and there is no discernible difference between the two solutions. Figure \ref{fig:GRINvsStair}(c) shows the GRIN  model solution for $\kappa = 10^{-5}$ - the localization has been suppressed entirely by the losses. Note however that the solution of the lossy staircase model shown in \ref{fig:GRINvsStair}(d) retains the localization feature and is essentially an attenuated form of the solution shown in \ref{fig:GRINvsStair}(b). Hence the staircase model is quantifiably different from the lossy version of the GRIN model. This prompts investigating whether a simple element can be incorporated into the GRIN model that captures both the lossless and lossy behaviours shown by the staircase model.

The element that makes both GRIN and staircase models qualitatively consistent with each other is the inclusion of a (single) discontinuity of size $\delta$ located at $z=\overline{z}$, see equation (\ref{eq:fullform}), as depicted in figure \ref{fig:GRIN}(a). Figure \ref{fig:PHwdelta} quantifies the sensitivity of this $\delta$-GRIN model with the size of $\delta$ for an angle of incidence $\theta = \pi/8$. Figure \ref{fig:PHwdelta}(a) shows the dependence of the size of the peak value of the localized field with the dimensionless layer width. The blue (stars) and green (diamonds) curves are the lossless and lossy ($\kappa = 10^{-3}$) cases respectively, as previously seen in figure \ref{fig:lossy}(a), and these act as bounds for the lossy $\delta$-GRIN model, where the value of $\delta=0.1$ dark yellow (squares), $\delta=0.05$ purple (triangles) and $\delta=0.01$ blue (circles). It can be seen that these results match smoothly to the lossless results for small values of $ka$, and for sufficiently small values of $ka$ are independent of  $\delta$. Figure \ref{fig:PHwdelta}(b) shows the dependence of the transmission coefficient as a function of $\delta$ (note, a value of $\delta=2$ is equivalent to step change between the bulk right- and left-handed media without a diffuse layer). The curves displayed are for different values of the dimensionless layer width, $ka = 1$ purple (triangles), $ka = 10^{-1}$ dark yellow (squares) and $ka = 10^{-2}$ green (circles). Thus the more diffuse the layer is, the less radiation is transmitted to the left-handed medium. 

\section{The simultaneous limit $\delta \to 0^+$ and $\kappa \to 0^+$}
We have shown that the GRIN model displays different behaviour according to whether $\kappa = 0$ or is finite, but that the latter case can be made consistent with the staircase model with the inclusion of a discontinuity $\delta$ in the refractive index profile, and this we have termed the $\delta$-GRIN model. There is clearly singular behaviour in the nature of the solutions obtained with these models as $\delta$ and $\kappa$ both tend to zero, which recovers the lossless GRIN model. This section discusses the reason for the discrepancy. 

We may define two classes of solution obtained in terms of whether or not there exists localization of the field within the layer, and the class that the solution adopts is different according to whether we take $\delta \to 0$ followed by $\kappa \to 0$  (no localization), or $\kappa \to 0$  followed by $\delta \to 0$ (localization). If $\delta \gg \kappa$ (in magnitude) as $\delta \to 0$, then the field will show pronounced localization features.
\begin{figure}[t!]
\center
\includegraphics[width=0.6\textwidth]{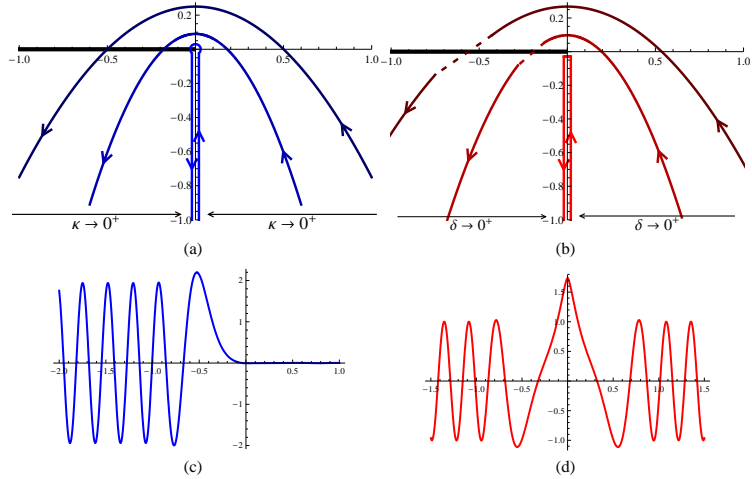}
\caption{(Colour Online) An illustration of the complex path for the third component of the U as the layer is traversed. The thick black line indicates the branch cut and the paths are for different losses. (a) shows the GRIN solution and that as $\kappa \to 0$ the path lines up to the negative imaginary axis whilst including the branch cut contribution. (b) shows that the introduction of the discontinuity causes the branch cut to be avoided (dashed line) leading to a lossless solution along the negative imaginary axis that does not include the branch cut. The real parts of the electric field across a lossless layer are shown in (c) and (d) for the limit of (a) and (b) respectively, using typical values for any variables used}
\label{fig:BC}
\end{figure}

The terminology of Leonhardt \cite{ConformalMapping} provides an alternative way to view the behaviour of the two models, depending on whether the branch cut of the U hypergeometric function is crossed. The quantity that determines this is the third argument of U, which is $\Lambda (z)  = i \gamma \Psi (z)$, see equation (\ref{eq:slelec}). In the lossless case the path of $\Lambda (z)$ causes the third argument of U to touch its branch cut at the origin, but not necessarily cross it, this gives rise to the two distinct solutions seen previously. The axes of figure \ref{fig:BC}(a-b) are the real and imaginary parts of $\Lambda(z)$ and the curves are the loci of $\Lambda (z)$. The branch cut of U is Re$(\Lambda (z)) < 0$ and Im$(\Lambda (z)) = 0$, which is shown by the thick black lines. Figure \ref{fig:BC}(a) is for the GRIN model, and the paths of $\Lambda (z)$ must cross the branch-cut, so that the solutions move to the next Riemann sheet, and give rise to the non-localized solutions as depicted in figure \ref{fig:BC}(c). Figure \ref{fig:BC}(b) is appropriate for the $\delta$-GRIN model, and the presence of a non-zero value of $\delta$ means that the locus of $\Lambda(z)$ does not cross the branch cut, but can `jump' across it with the solution remaining on the same Riemann sheet - hence the localization form of the solution is obtained as shown in Figure \ref{fig:BC}(d).

\section{Conclusion and discussion}
In this paper we have introduced a model to account for smooth changes of permittivity and permeability across a diffuse boundary between bulk right- and left-handed metamaterials. A full-wave, exact analytical solution to this problem leads to a strong localisation of the field in the transition region whilst being fully transmissive. In the lossy case the strong localisation is removed and a reflected wave exists. Consideration of another analytical model leads to the inclusion of a discontinuity in the refractive index profile which restores qualitatively the features of the lossless case. In all cases this paper has been able to analytically quantify the reflected and transmitted wave properties.

The reason for the localization in the layer is the constructive interference of evanescent modes that are stimulated whenever $\left| n \right| < \sin \theta$ as illustrated in figure \ref{fig:lossy}. It should be stressed that these modes are not a conventional plasmon mode which is generated by a discontinuous change in the refractive index \cite{Ruppin}. In contrast the diffuse layer causes the coherent addition of a plasmon and an anti-plasmon \cite{AntiPlasmon} throughout the \emph{volume} of the metamaterial for which $\left| n \right| < \sin \theta$.

Although not detailed here, equation (\ref{eq:slprofile}) can also be used to model changes between two right-handed media or indeed two left-handed media. Altogether this method can model a diffuse boundary between any combination of left- and right-handed media, with or without losses in either medium.

Another practical application of the model is the extension of the graded index approach to modelling surface roughness as a graded index change, e.g. \cite{Lekner}, to include magnetic and left-handed materials. Also, since the solution contains the polarization-state of the wave the approach can be used to investigate, for example, the emission polarization effects of infra-red radiation, e.g. \cite{IREmission}, from left-handed media. Other applications include further polarization effects, such as the Brewster angle, and to analyse how superlensing properties are sensitive to the polarization state and by roughness of the lens surfaces, and these will be treated elsewhere.

During the period that this paper was being refereed, a paper \cite{otherpaper} has studied a similar model to that presented here but with different emphasis. 

\section{Acknowledgments}
PI is funded by the UK's Engineering and Physical Sciences Research Council (EPSRC).

\appendix

\section{Exponential Layer dependance} \label{cp:exp}
The other most useful profile, is that of an exponential dependence across the layer:
\begin{equation}
\mu = \exp (m \, z), \quad \epsilon = \eta \, \mu + A
\end{equation}
with the solution
\small
\begin{equation}
\label{eq:expgen}
E(z) = \left( \frac{\mathcal{Z}(z) \, c}{i \, \eta^{1/2} \, \omega} \right)^{\frac{\chi}{2}} \exp \left( - \mathcal{Z}(z) \right) \left( \alpha F(z) + \beta G(z) \right)
\end{equation}
\normalsize
where 
\small
\begin{equation*}
F(z) = \mbox{U} \! \left( \frac{\chi + \frac{i \,A\,c}{m\,{\eta^{1/2}}\,\omega }}{2}, \chi, 2 \, \mathcal{Z}(z) \right) ,
\end{equation*}
\begin{equation*}
G(z) = \mbox{L} \! \left( \frac{-\chi - \frac{i \,A\,c}{m\,{\eta^{1/2}}\,\omega }}{2},\chi-1, 2 \, \mathcal{Z}(z) \right) ,
\end{equation*}
\begin{equation*}
\chi = 1 + {\left( 1 + \frac{4\,k_x^2}{m^2} \right)^{1/2} }, \quad \mathcal{Z}(z) = \frac{i \exp(m \, z) \eta^{1/2} \omega}{c \, m} ,
\end{equation*}
\normalsize
$\mbox{U}$ and $\mbox{L}$ being the Hypergeometric U and Laguerre L functions respectively and $\alpha$ and $\beta$ are constants of integration. In the case that $A=0$, i.e. that $\epsilon(z)=\eta \, \mu(z) = \eta \, \exp (m z)$ then the U and L functions simplify to the Bessel J and Y functions. The most useful applications of this model have $\mu$ and $\epsilon$ continuous across $z = z_1$ and $z = z_2$, c.f. section \ref{cp:grin}. This allows for a large number of possible changes between two RHM, but for a change from a RHM to a LHM $m$ must be complex and necessarily introduces losses into the model which are largely uncontrollable. Still this model exists and may be better suited to some applications. 

\bibliographystyle{unsrt}

\end{document}